\documentclass[a4paper]{jpconf}

\usepackage{graphicx}

\usepackage{hyperref}

\usepackage[sort&compress,numbers]{natbib}

\usepackage{amsmath}
\usepackage{amssymb}
\usepackage{bm}

\newcommand{\maestroex}{{\sffamily MAESTROeX}}
\newcommand{\castro}{{\sffamily Castro}}

\newcommand{\amrex}{{\sffamily AMReX}}

\newcommand{\Uc}{{\,\bm{\mathcal{U}}}}

\newcommand{\Advs}[1]{\boldsymbol{\mathcal{A}} \left(#1\right)}

\newcommand{\avg}[1]{{\left \langle #1 \right \rangle}}
\newcommand{\Rbs}[1]{{\bf R} \left ( #1 \right )}

\usepackage{color}
\begin{document}

\title{The Castro AMR Simulation Code: Current and Future Developments}

\author{M. Zingale$^1$,
        A.~S.~Almgren$^2$,
        M.~Barrios Sazo$^1$,
        J.~B. Bell$^4$,
        K.~Eiden$^1$,
        A.~Harpole$^1$,
        M.~P. Katz$^3$,
        A.~J. Nonaka$^2$,
        D.~E. Willcox$^2$, and
        W. Zhang$^2$}

\address{$^1$Department of Physics and Astronomy, Stony Brook
  University, Stony Brook, NY 11794-3800 USA}

\address{$^2$Center for Computational Sciences and Engineering,
  Lawrence Berkeley National Lab, Berkeley, CA 94720 USA}

\address{$^3$NVIDIA Corporation, 2788 San Tomas Expressway,
  Santa Clara, CA, 95051 USA}

\ead{michael.zingale@stonybrook.edu}

\begin{abstract}
We describe recent developments to the \castro\ astrophysics
simulation code, focusing on new features that enable our simulations
of X-ray bursts.  Two highlights of \castro's ongoing development are
the new integration technique to couple hydrodynamics and reactions to
high order and GPU offloading.  We discuss how these features will
help offset some of the computational expense in X-ray burst models.
\end{abstract}

\section{Introduction}

The \castro\ astrophysical simulation code~\cite{castro} is designed
for modeling problems in nuclear astrophysics, with the ability to
accurately capture the interplay between hydrodynamics, reactions,
gravity, and radiation in stars with complex equations of state.
Since \castro\ was first developed, there have been a number of
enhancements to the code base, expanding its applicability to a new
range of scientific problems.  Throughout this development,
\castro\ has strived to perform well on modern supercomputers,
adapting to trends in hardware and programming models, while
maintaining portability across architectures.

\castro\ has been applied to models of Type Ia supernovae,
core-collapse supernovae, pair-instability supernovae, exoplanet
atmospheres, and most recently X-ray bursts (see, e.g.,
\cite{castro-ccsne,castro-pairinstability,polin:2019,wdmergerI} for
some example science applications).  A common challenge in modeling
these events is the range of length and timescales involved.  To
capture length scales, \castro\ uses adaptive mesh refinement (AMR), through
the \amrex\ library~\cite{amrex_joss}.  This allows us to focus the
computational effort on regions where burning or the flow is
important.  To address the range of timescales involved in
astrophysical explosions, we have developed a low Mach number
hydrodynamics code, \maestroex~\cite{maestroex}, built on the same
framework as \castro, that can model the subsonic convection that
often precedes explosive events.  Both codes are open source and
freely available on
github\footnote{\url{https://github.com/AMReX-Astro/}}.

The most active development presently focuses on new methods of time
integration, with better coupling of physical processes, and GPU
performance.  In these next sections we discuss these features and
their impact on our ongoing X-ray burst simulations.

\section{Modeling Reactive Flow}

In modeling stellar explosions, it is often the case that the
hydrodynamics method and the nuclear reaction network require
different timesteps in order to produce stable and accurate models.
For the stiff nuclear reactions we encounter in stars, implicit ODE
methods are often used together with operator splitting to evolve the
nuclear reactions separately from the hydrodynamics.  In the limit of
small timesteps, the two processes are well-coupled together, but
these conditions are not always met in simulations.  The recent focus
in \castro\ has been on high-fidelity simulations of reactive flow.
In \cite{castro:sdc} we introduced a new time integration strategy,
spectral deferred corrections (SDC), that eliminates the coupling
error introduced by commonly used operator splitting techniques (see
the discussion in \cite{astronum:2018} for a graphical illustration of
splitting error).

The SDC algorithm used in \castro\ follows the ideas of
\cite{dutt:2000,minion:2003}, and uses low order explicit advection
and implicit reaction updates in a correction equation that when
applied iteratively achieves high-order time-accuracy.  In an operator
split implementation, reaction and burning proceed without knowing
about the other process.  In contrast, the update in the SDC
formulation explicitly couples the processes together.  In \castro\ we
consider reactions and hydrodynamics, and write our conservative
system as:
\begin{equation}
\Uc_t = \Advs{\Uc} + \Rbs{\Uc}
\end{equation}
where $\Uc$ is the vector of conserved quantities, $\Advs{\Uc}$ is the
advective term (the divergence of the hydrodynamic fluxes along with
hydrodynamic sources), and $\Rbs{\Uc}$ are the reactive source terms.
The update is done at specific temporal nodes, the number of locations
of which are picked to give the desired temporal accuracy.  The update
from one time node $m$ to $m+1$ appears as:
    \begin{align}
      \label{eq:sdc:general}
      \avg{\Uc}^{{m+1},(k+1)} = \avg{\Uc}^{m,(k+1)}
            &+ \delta t_m \left [ \avg{\Advs{\Uc}}^{m,(k+1)} - \avg{\Advs{\Uc}}^{m,(k)} \right ] \nonumber \\
            &+ \delta t_m \left [ \avg{\Rbs{\Uc}}^{{m+1},(k+1)} - \avg{\Rbs{\Uc}}^{{m+1},(k)} \right ] \nonumber \\
            &+ \int_{t^m}^{t^{m+1}} dt  \left (\avg{\Advs{\Uc}}^{(k)} + \avg{\Rbs{\Uc}}^{(k)}\right)
   \end{align}
We use a finite-volume formalism, so $\avg{\Uc}$ is the spatial average of the
conserved state in a zone, $\avg{\Advs{\Uc}}^m$ is the average
advective term in a zone at time $t^m$, and $\avg{\Rbs{\Uc}}^{m+1}$
the average of the reactive source at time $t^{m+1}$.  The second
superscript on each term, $(k)$ or $(k+1)$, is the iteration counter.  This
update is an implicit equation for the new state,
$\avg{\Uc}^{{m+1},(k+1)}$.  The last term in the update is an integral
over the sources constructed from the previous iteration's values at
each time node.  Each iteration of the SDC method increases the
temporal order of accuracy by one, up to the order of accuracy with
which the integral is constructed.  

\castro\ implements both a second-order method (using a trapezoid rule
for the integral) and fourth-order method in space and time (using a
Simpsons rule for the integral and the spatial reconstruction of
\cite{mccorquodalecolella}).  We demonstrated that \castro\ achieves
the expected convergence on a wide variety of problems in
\cite{castro:sdc}.  At the moment, the method is limited to single
levels, but work is underway to extend this methodology to AMR.  For
problems where burning is important and can dominate the computational
expense, we hope this new SDC method will become the preferred
integration technique in the future.

\section{Performance Portability}

\begin{figure}[t]
\centering
\includegraphics[width=0.8\linewidth]{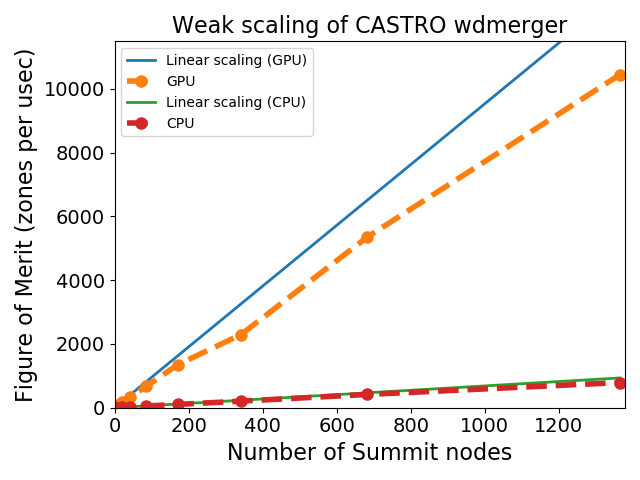}
\caption{\label{fig:gpu} \castro\ weak scaling on the OLCF Summit
  machine for the white dwarf merger problem.  On each node we use
  either 42 IBM Power 9 CPU cores or 6 NVIDIA Volta GPUs.  Ideal
  scaling is shown for both cases as the solid line.}
\end{figure}

The original approach to parallelism in \castro\ was MPI + OpenMP,
with scaling on manycore architectures achieved using a tiling approach
to OpenMP~\cite{tiling,astronum:2017}.  The general approach is to
have MPI distribute the AMR grids across nodes and have OpenMP threads
work on regions of each grid (tiles) by passing the box describing the
tile into the computational kernel.  More recently we've ported
\castro\ to GPUs using CUDA, using the same computational kernels as
the MPI + OpenMP version.  When run on GPUs, each CUDA thread handles
the update of a single zone, simply by passing that zone index into
the computational kernel.  This reuse keeps the code base
manageable---we don't need separate kernels for each
architecture---while allowing us to take advantage of current and next
generation supercomputers.  All of the solvers needed to run our core
science problems run on GPUs: the main unsplit PPM hydrodynamics
scheme~\cite{ppm,millercolella:2002}, self-gravity via multigrid with
isolated boundary conditions, thermal diffusion, and nuclear
reactions.  Our approach is to put the data on GPUs and then run all
of the kernels on the GPUs, minimizing data movement.  This has given
us enormous speed-ups.  Figure~\ref{fig:gpu} shows weak scaling on the
OLCF Summit machine for a WD merger problem~\cite{wdmergerI}.  When
using CPUs, we use 42 IBM Power 9 cores per node, while when using
GPUs we use 6 NVIDIA Volta GPUs per node.  We see that the code
performance is about $10\times$ higher per node when using GPUs, while
both show excellent weak scaling.

\section{Example Application: X-ray bursts}

Both the new SDC method and GPU offloading are important to our
science goals, in particular the problem of modeling a flame spreading
across the neutron star during an X-ray burst (XRB).  We discussed the
computational challenges of modeling these events in
\cite{astronum:2018}.  In short: the largest scale we need to model is
the size of the neutron star, or at least several times the scale at
which rotation and the lateral pressure gradient balance (the Rossby
length).  The smallest scale we need to model is dictated by
accurately capturing the burning---we are interested in both the
energy release and nucleosynthesis.  Our first set of two-dimensional simulations
(\cite{astronum:2018}, Eiden et al.\ in preparation) used two physical approximations to
make the problem more tractable.  First, we used a higher rotation
rate than expected in order to reduce the Rossby length, allowing us
to model a smaller region of that star.  Second, we artificially
boosted the speed of the flame to reduce the duration we need to
model.  Our next set of calculations will relax these approximations.

With the new fourth-order accurate reactive hydro solver, we believe
we can capture the dynamics of the spreading flame accurately with one
less refinement level.  This will allow us to expand to larger domains
while keeping the memory demands reasonable.  Full science simulations
will begin once we port the SDC framework to AMR and generalize the
fourth order solver to axisymmetric geometries.  Proof-of-concept single
level runs are running now, but the lack of AMR makes them very
expensive.

The large increase in performance of the code when run on GPUs enables
us to do away with the flame boosting.  All of the physics solvers
needed for the XRB simulations run on the GPUs: the hydrodynamics,
explicit diffusion, realistic equation of state and conductivities,
and the 13-isotope He burning network we are using.  Preliminary
estimates show that we can expect a speed-up of at least $5\times$ for the OLCF Summit
nodes (using GPUs) over the NERSC Cori Haswell nodes (using CPUs).
Work is underway to further optimize the GPU code,
especially the reaction networks.

\section{Future Developments}

Since its inception \castro\ has undergone steady development to
enable new science investigations and take advantage of new
supercomputer architectures.  The development of SDC, summarized here,
continues, with the current focus on enabling our XRB studies.  We are
also investigating astrophysical detonations with SDC to understand
how well the method works with nuclear statistical equilibrium.  For
the detonation work, we are exploring different quadrature schemes for
the integral in Equation~\ref{eq:sdc:general} that are more amenable
to highly-stiff reaction processes.  We are also working on extending
the SDC integration methodology to adaptive mesh refinement with
subcycling in time.

To complement the existing suite of hydrodynamics solver in \castro,
an MHD solver is under development and expected to be merged into the
main branch soon.  We will use the experiences learned with the
\castro\ hydrodynamics solver to port this solver to GPUs and fit into
the SDC framework.

Altogether, these developments will allow us to begin our first
three-dimensional realistic XRB calculations by the end of the year,
with parameter studies to follow.  These calculations will help us
understand the nucleosynthesis during XRBs and its impact on
observables along with the interpretation of lightcurves and neutron star
structure.

\ack The work at Stony Brook was supported by DOE/Office of Nuclear
Physics grant DE-FG02-87ER40317 and contract 7418390 with Lawrence
Berkeley National Laboratory as part of the Exascale Compute Project
ExaStar collaboration.  This research was supported by the Exascale
Computing Project (17-SC-20-SC), a collaborative effort of the
U.S. Department of Energy Office of Science and the National Nuclear
Security Administration.  The work at LBNL was supported by the DOE
Office of Advanced Scientific Computing Research under Contract No,
DE-AC02-05CH11231.  An award of computer time was provided by the
Innovative and Novel Computational Impact on Theory and Experiment
(INCITE) program. This research used resources of the Oak Ridge
Leadership Computing Facility at the Oak Ridge National Laboratory,
which is supported by the Office of Science of the U.S. Department of
Energy under Contract No.\ DE-AC05-00OR22725.  This research used
resources of the National Energy Research Scientific Computing Center,
which is supported by the Office of Science of the U.S. Department of
Energy under Contract No.\ DE-AC02-05CH11231.  Visualizations were
done using yt~\cite{yt}.  This research has made use of NASA's
Astrophysics Data System Bibliographic Services.


\providecommand{\newblock}{}

\end{document}